\documentstyle[aps,preprint,psfig]{revtex}
\tightenlines
\begin{document}
\draft
\preprint{}

\title{Superconductivity in Mesoscopic Metal Particles: 
The Role of Degeneracy}

\author{H. Boyaci, Z. Gedik and I. O. Kulik\footnote{Also at: B. Verkin 
Institute for Low Temperature Physics and Engineering, Natl. Acad. Sci. 
of Ukraine, Kharkov 310164, Ukraine}}

\address{Department of Physics, Bilkent University, 
Bilkent 06533 Ankara, Turkey}

\date{\today}
\maketitle
\begin{abstract}

Recently, it has been possible to construct single-electron transistors to
study electronic properties, including superconductivity, in metallic 
grains of nanometer size. Among several theoretical results are suppression 
of superconductivity with decreasing grain size and parity effect, i.e. 
dependence on the parity of the number of electrons on the grain. We study 
how these results are affected by degeneracy of energy levels. 
In addition to the 
time-reversal symmetry, for certain energy spectra and more 
generally for lattice symmetries, energy levels are strongly degenerate near 
the Fermi energy. For a parabolic dispersion, degeneracy $d$ is of 
the order of $k_FL$ while the typical distance between the levels is of the 
order of $\epsilon_F/(k_FL)^2$ where $k_F$ and $\epsilon_F$ are 
the Fermi wave vector and energy, respectively, and $L$ is the 
particle size. First, using an exact solution method 
for BCS Hamiltonian with finite number of energy levels, 
for the well studied non-degenerate case
we find a new feature. In that case, parity effect exhibits  
a minimum instead 
of a monotonic behavior.
For $d$-fold degenerate states we find that the ratio of 
two successive parity effect parameters $\Delta_p$ is nearly 
$1+1/d$. Our numerical solutions for the exact ground state energy of 
negative $U$ Hubbard model on a cubic cluster also give very similar
results. Hence we conclude that parity effect is a general property
of small Fermi systems with attractive interaction, and
it is closely related to degeneracy of energy levels.

\end{abstract}


\narrowtext

\section{Introduction}
Recently, small metallic grains, with sizes down to 5-10 nm, became
available for experimental study\cite{tinkham,black,ralph}. Superconducting
transition in such grains manifests itself through the {\em parity effect},
namely dependence of electronic properties of the grain on whether the number of
electrons in the grain is even or odd. Tinkham, Hergenrother and
Lu\cite{tinkham} showed that in a single-electron tunneling (SET)
transistor\cite{averin} with a superconducting island, conductance channels
open below a certain temperature $T^*<T_c$ only for an odd number of
electrons on the superconducting grain, whereas above $T^*$ they open
symmetrically at odd and even number of electrons.  This means that
energy of the state with odd number of electrons shifts up by an amount
$\Delta_p$\cite{matveev}, so called {\em parity effect parameter}
\begin{equation}
\Delta_p=E_0^{(2N+1)}-\frac{1}{2}(E_0^{(2N)}+E_0^{(2N+2)}).
\end{equation}

Several authors \cite{vondelft,smith,mastellone,berger} 
developed theories attempting 
to calculate properties of superconducting state of the 
grain with discrete electronic
levels. It was assumed that the level degeneracy is totally removed, and the
only parameter distinguishing the small grain from the bulk sample is
the ratio of the average level spacing $\delta$ in a sample to the
bulk superconducting gap $\Delta$, $\delta/\Delta$. 

There are two main motivations for studying the effect of degeneracy.
First, small superconducting particles can in principle
be prepared in a perfect symmetric shape \cite{landman} 
in which case the mesoscopic 
superconductivity parity effect will show
novel features due to level degeneracy. The second and
more important is the necessity of distinguishing between 
the level spacing due to size effect 
$\delta_1 \sim \hbar v_F/L$, and the average
level spacing $\delta_2 \sim \epsilon_F/N$ which is much smaller 
than $\delta_1$
($N$ is total 
number of conduction electrons in metallic grain). The ratio 
$\delta_1/\delta_2$ is of the order
$N^{2/3}$, which is very large for a typical mesoscopic particle of
size $0.1\mu\rm m$ corresponding to $N\sim10^6$. 
The realistic energy structure of small grain is
therefore not equidistant spectrum with level spacing
$\delta_1$, adopted in most papers on
mesoscopic superconductivity, but rather peaks in 
the density of states with characteristic distance $\delta_1$
accompanied by finer structure
with smaller energy separation $\delta_2$.
To understand which effects such finer 
structure may have on superconductivity in mesoscopic particles,
we adopt a model of $\delta_1$-spaced 
levels with the degeneracy $d$ (in the rest of our work 
we drop the index and denote level spacing by $\delta$). 
Calculation of superconducting 
condensation energy \cite{abrikosov} shows that energy binding per particle is
{\em not} $\Delta$, as may be expected from a naive picture of
particle binding resulting in energy decrease $2\Delta$, but rather of order 
$\Delta^2/\epsilon_F$. Adding extra two electrons to superconductor
decreases energy by $2\Delta$ but at the same time provides 
a (small) shift of the chemical potential in such a way that 
the net energy change per particle 
is only $\Delta \times \Delta/\epsilon_F$. A realistic calculation of the
parity effect must treat system of interacting electrons 
self-consistently as it is done in the standard BCS theory of bulk 
superconductors \cite{abrikosov}.

In the theory of superconductivity of Bardeen, Cooper and
Schrieffer\cite{bardeen}, the coupling Hamiltonian is introduced in the form 
\begin{equation}
H_{int}=
g\sum_{{\bf k}{\bf k}'}
c_{{\bf k}\uparrow}^\dagger 
c_{-{\bf k}\downarrow}^\dagger 
c_{-{\bf k}'\downarrow}
c_{{\bf k}'\uparrow}
\label{bcs}
\end{equation}
where operator $c_{{\bf k}\sigma}^\dagger$ creates an electron in a state
with momentum ${\bf k}$ and spin projection $\sigma=\uparrow,\downarrow$.
${\bf k}$ and $-{\bf k}$ states are selected in expense of all other states
(such as
${\bf k}$ and ${\bf q}$ with ${\bf q}\neq-{\bf k}$), because only these two
time-reversed states lead to singularity which results in Cooper instability
identified by logarithmic divergence of the scattering amplitude
$\ln(2\hbar\omega_D/\epsilon)$ as $\epsilon\rightarrow 0$ \cite{abrikosov}. 
Pairing of non-time reversed states can be treated
perturbatively, or can be included as a renormalization in the Fermi liquid
picture. 

Assume that the single electron states $\psi_\eta$ in metal are decomposed
as
\begin{equation}
\psi_\eta=\sum_{\bf k}a_{\bf k}^\eta\phi_{\bf k}
\end{equation}
where $\phi_{\bf k}$ are plane waves. We are going to denote the time reversal
state corresponding to $\psi_\eta$ by $\psi_{\bar{\eta}}$ which is given by the
complex conjugate of $\psi_\eta$. 
In general $\psi_\eta$
and $\psi_{\bar{\eta}}$ are different states.
Let us introduce the operators
\begin{equation}
c_{\eta\sigma}^\dagger=\sum_{\bf k}a_{\bf k}^\eta c_{{\bf
k}\sigma}^\dagger.
\end{equation}
These are Fermi operators since $c_{{\bf k}\sigma}^\dagger$ and 
$c_{{\bf k}\sigma}$ obey the Fermi statistics. Let us consider the sum
$A^\dagger=\sum_\eta c_{\eta\uparrow}^\dagger c_{\bar{\eta}\downarrow}^\dagger$,
which can be represented as
\begin{equation}
A^\dagger=\sum_{{\bf k}{\bf k}'}(\sum_\eta a_{\bf k}^\eta a_{{\bf
k}'}^{\bar{\eta}})c_{{\bf k}\uparrow}^\dagger c_{{\bf k}'\downarrow}^\dagger.
\end{equation}
If we form an interaction Hamiltonian using $A^\dagger$ and $A$ as 
$\tilde{H}_{int}=gA^\dagger A$, $\tilde{H}$ contains a
singular part which is identical to Eq. \ref{bcs} because of the identities 
$a_{\bf k}^\eta = (a_{-{\bf k}}^{\bar{\eta}})^*$ and 
$\sum_\eta|a_{\bf k}^\eta|^2=1$ (for real $\psi_\eta$, $|\eta\sigma\rangle$ and
$|\bar{\eta}\sigma\rangle$ turn out to be the same state and 
if there is no further degeneracy due to other symmetries $H_{int}$ reduces to 
the ``toy model" of superconductivity 
which has been studied extensively in the literature \cite{braun}).
Therefore, the interaction Hamiltonian in a grain can in general be written
as 
\begin{equation}
H_{int}=g\sum_{\eta\eta'}c_{\eta\uparrow}^\dagger
c_{\bar{\eta}\downarrow}^\dagger
c_{\bar{\eta}'\downarrow}
c_{\eta'\uparrow}.
\label{int}
\end{equation}
We will further split index $\eta$ into $j,\alpha$ where $j$ denotes the
energy levels and $\alpha$ denotes the
degenerate states for this level. 
The only 
symmetry expected to hold in an irregularly shaped ultrasmall grain is the 
time-reversal symmetry which may lead to double degeneracy for the energy
levels, for example current carrying states. On the other hand, 
for certain energy spectra and more 
generally for lattice symmetries, energy levels are strongly degenerate near 
the Fermi energy. For very small particles, splitting due to weak disorder or
asymmetry is much smaller than the splitting 
due to size effect. In this case the system can be modeled as almost 
degenerate with two dimensionless parameters governing superconducting 
transition, namely $\delta/\Delta$, the ratio of 
level spacing to bulk gap, and $d$, the degeneracy of split levels.

\section{Parity Effect for Degenerate Levels}
In our study,
for simplicity, we assume all energy levels are $d$-fold degenerate
and levels are equally spaced with level spacing being equal to $\delta$.
We start with the following pair interaction Hamiltonian 
which is properly modified to
take the degeneracy into account 
\begin{equation}
H=\delta \sum_{j\alpha\sigma}jc_{j\alpha\sigma}^\dagger c_{j\alpha\sigma}
-\frac{\lambda\delta}{d}\sum_{j,j' \in S;\alpha \alpha'}
c_{j\alpha\uparrow}^\dagger 
c_{j\bar{\alpha}\downarrow}^\dagger 
c_{j'\bar{\alpha}'\downarrow}
c_{j'\alpha'\uparrow},
\end{equation}
where second sum is over the set of levels $S$ which are lying
within the $2\hbar\omega_D$ shell centered at Fermi level, that is
$j,j^\prime\in S=\{-n_c,\ldots,n_c\}$ with $n_c$ being equal to 
integer part of $\hbar\omega_D/\delta$, and $\alpha=1,\ldots,d$
runs over all the degenerate states for a given energy level.
Here, $|j'\bar{\alpha}'\downarrow\rangle$ and $|j'\alpha'\uparrow\rangle$
are time reversed states for which the matrix elements are much larger then
all others\cite{anderson}. For example, for a grain where eigenstates are
labeled by crystal momentum ${\bf k}$, the two states are $|{\bf
k}\uparrow\rangle$ and $|-{\bf k}\downarrow\rangle$. Note that there is
another similar but different pair formed by $|{\bf k}\downarrow\rangle$ 
and $|-{\bf k}\uparrow\rangle$. In usual BCS problem, since there is a 
summation over ${\bf k}$, both pairs are properly taken into account.
On the other hand, when we sum over energy levels rather than individual 
states we
must be careful in including both pairs. However, the model without double
degeneracy can still be used to describe the superconductivity in systems 
with real wave functions \cite{braun}. 

Next, we introduce the boson operators
\begin{equation}
n_{j\alpha}=\frac{1}{2}(c_{j\alpha\uparrow}^\dagger
c_{j\alpha\uparrow}+ 
c_{j\bar{\alpha}\downarrow}^\dagger 
c_{j\bar{\alpha}\downarrow})
\end{equation}
and 
\begin{equation}
b_{j\alpha}=c_{j\bar\alpha\downarrow}
c_{j\alpha\uparrow}.
\end{equation}
It is easy to verify that $b_{j\alpha}^\dagger$ and $b_{j\alpha}$
satisfy ``boson" commutation relations
\begin{equation}
[b_{j\alpha},b_{j^\prime\alpha^\prime}^\dagger]=
\delta_{jj'}\delta_{\alpha\alpha'}(1-2n_{j\alpha}).
\end{equation}
Presence of the number operator on the right hand side is direct consequence
of the Pauli principle. With these new operators,
the pairing Hamiltonian can be rewritten as
\begin{equation}
H=2\delta\sum_{j\alpha}jn_{j\alpha}
-\frac{\lambda\delta}{d}\sum_{j,j'\in S;\alpha \alpha'}
b_{j\alpha}^\dagger b_{j'\alpha'}
\label{bose}
\end{equation}
This Hamiltonian can be split into two parts as $H=H_1+H_{eff}$
where
\begin{eqnarray}
H_1=2 \delta \sum_{j\not\in S;\alpha} j n_{j\alpha}  \nonumber\\
H_{eff}= 2\delta \sum_{j\in S; \alpha} j n_{j\alpha} -
        \frac{\lambda \delta}{d}\sum_{j,j'\in S;\alpha,\alpha'}
        b_{j\alpha}^\dagger b_{j'\alpha'},
\end{eqnarray}
and the sum in $H_1$ is over non-interacting particles while
in
$H_{eff}$ both of the sums are over interacting particles. 
Interacting particles are those which 
occupy levels inside $2\hbar\omega_D$ shell about the Fermi level except the 
singly occupied level. Since the interaction is only between pairs,
this singly occupied level is removed form the set $S$. This is 
the so called ``blocking effect".
As $H_1$ and $H_{eff}$ commute, eigenstates of $H$ are product of eigenstates
of $H_1$ and $H_{eff}$, and eigenvalues of $H$ are sums of corresponding 
eigenvalues of $H_1$ and $H_{eff}$. $H_1$ simply represents 
non-interacting particles in a potential, thus determining the ground state 
energy of $H$ reduces to solving for $H_{eff}$. 

The effective Hamiltonian $H_{eff}$ has been studied with exact
diagonalization methods by Mastellone, Falci, and Fazio (MFF)\cite{mastellone}, 
and by Berger and Halperin (BH) \cite{berger} for the non-degenerate case 
and with relatively small number of levels taken into account ($n_c<15$).
However, there exists an exact solution for $H_{eff}$ due to 
Richardson and Sherman (RS) \cite{richardson}. This long forgotten 
exact solution has been re-introduced to condensed matter community 
by Braun and von Delft \cite{vondelft2} in the context of ultrasmall 
superconducting grains. Here, we briefly review 
the Richardson-Sherman solution.
First,
$H_{eff}$ is treated without the Pauli exclusion principle,
that is 
hardcore Bose particles are treated as normal Bose particles.
After diagonalizing that Hamiltonian for normal bosons
by a canonical transformation, 
the following boson 
wave function is obtained
\begin{equation}
\psi_b(j_1\ldots j_N)=\sum_p P \{ \prod_{k=1}^N \frac{1}{2\delta j_k-E_{p_k}} \},
\label{wave}
\end{equation}
where $N$ is the total number of bosons, and
in $\sum_p$, $P$ means summing over $N!$ permutations of indices
$p_1,\ldots,p_N$, and the corresponding energy is
\begin{equation}
E=E_{p_1}+\ldots+E_{p_N}.
\label{groundenergy}
\end{equation}
Next, Pauli exclusion principle is imposed with the condition that $\psi=0$ 
if any two $j_i$ are not distinct. This restriction is introduced by writing 
the wave function as 
\begin{equation}
\psi(j_1 \ldots j_N)=\theta(j_1 \ldots j_N) \phi(j_1 \ldots j_N)
\label{wavefunction}
\end{equation}
where $\theta(j_1 \ldots j_N)$ is equal to 1 if all $j_i$'s are distinct, and
is equal to 0 otherwise. 
Then, after some 
calculation an effective Schr\"{o}dinger equation for $\phi$ is obtained 
\cite{richardson}
\begin{equation}
(2\delta j_1+\ldots+2\delta j_N-E) \phi(j_1\ldots j_N)
-\frac{\lambda \delta}{d} \sum_i^N\sum_{j\ne j_i}(1-\sum_{k \ne i}^N\delta_{ik})\phi(j_1 \ldots 
j_{i-1},j,j_{i+1},\ldots j_N)=0.
\label{eqforwafu}
\end{equation}
For a multiply degenerate single level, ground state energy is given by
\begin{equation}
E_0=2 N \epsilon -\frac{\lambda \delta}{d} N (d-N+1),
\label{single}
\end{equation}
where $d$ is the pair degeneracy of the level. 
Next we focus on degenerate levels in the $2\hbar\omega_D$ 
interacting shell leading to a set of coupled equations
for $N$ pairs and $M$ levels which are $d$-fold degenerate:
\begin{equation}
1+2\frac{\lambda\delta}{d}\sum_{p\neq q}^N\frac{1}{E_p-E_q}
-\frac{\lambda\delta}{d}\sum_\alpha\sum_{j=-n_c}^{n_c}
\frac{|2\langle n_{j\alpha}\rangle-1|}{2\delta j-E_q}=0; \,\, q=1\ldots N
\label{rs}
\end{equation}
where
\begin{equation}
E_{p}\ne E_{q}, \,\, \mbox{for} \,\, p\ne q.
\label{restriction}
\end{equation}
The ground state wave function is given by (\ref{wave}) up to a 
normalization constant
and the corresponding ground state energy is calculated from (\ref{groundenergy}).

At some values of $\lambda\delta/d$, depending upon the state under
consideration, Eq. \ref{rs} has
singularities. At a singularity the restriction (\ref{restriction})
is not satisfied. However, the domain of validity of the solution 
can be extended by letting $E_q$ to be complex. Complex roots
$E_q$ occur in complex conjugate pairs
\begin{eqnarray}
E_{2\gamma}=x_\gamma+iy_\gamma, \,\,
E_{2\gamma-1}=x_\gamma-iy_\gamma, \nonumber\\
\gamma=1,\ldots,N/2,
\label{complexroots}
\end{eqnarray} 
where $x_\gamma$ is real and $y_\gamma$ can be either real or pure imaginary. 
It turns out that if $N$ is not even, one of the roots remains real for all $\lambda$.
This form of $E_q$ preserves the reality of the ground state energy $E$
(Eq. \ref{groundenergy}), and also the reality of the ground state 
wave function \cite{richardson}. At a singular point, no more than two 
roots can be equal, and these two equal pair energies are both
$2\delta j_i$ for some $j_i$. The desired roots of Eq. \ref{rs} 
for the ground state should satisfy
\begin{equation}
\lim_{\lambda\rightarrow 0} E_q=2\left(\left[\frac{q-1}{d}\right]+1\right)\delta, \,\, q=1,\ldots,N,
\label{conditions}
\end{equation}
that is, in the non-interacting system the lowest $N$ levels are occupied
by $N$ pairs while levels from $N+1$ to $M$ are 
unoccupied ($[\ldots]$ denotes integer part of the division). 
We solve 
Eq. \ref{rs} by a globally convergent Newton-Raphson method \cite{recipies}
for complex $E_q$ (\ref{complexroots}) 
with the conditions implied by Eq. \ref{conditions}.

For the non-degenerate case singularities of Eq. \ref{rs} can be 
removed by a suitable change of variables \cite{richardson2}.
However it is not possible to generalize that method to degenerate case.
In this case roots are complex for 
all values of $\lambda$ (if $N$ is odd, one root remains 
real for all $\lambda$). Fig. \ref{roots} shows typical behavior 
of roots for the double degenerate case (see Richardson \cite{richardson2} 
for the behavior of roots in the non-degenerate case).

The ground state
energy $E_0^{(n)}$ for a given number of electrons $n$ 
allows us to calculate 
the parity effect parameters
\begin{equation}
\Delta_P^{(i)}=|E_0^{(2n_cd+i-1)}
-\frac{1}{2}(E_0^{(2n_cd+i-2)}
+E_0^{(2n_cd+i)})|
\label{parity}
\end{equation}
where $i=1,\ldots,2d$. Here by $n$ we mean the number of electrons
in the thin shell around the Fermi level participating the pairing 
interaction. We assume that the shell is composed of $n_c$ levels below and 
above the Fermi energy. Note that with this definition $E_0^{(n)}$ can take
$2d$ different values and that is why $i$ index running from 1 to $2d$ 
exhausts all possible values of the right hand side.

In principle, to study the effect of finite energy level spacing,
we should fix the Debye frequency $\omega_D$, which is assumed
to be less affected by the boundary conditions, and change the
number of levels $n_c$. We estimate that typical $n_c$ values lie in the range
of $\hbar \omega_D /\delta \sim 50-2000$, and for $\delta\sim\Delta$,
we have $n_c\sim 100$. Numerical solution of
Eq. \ref{rs} becomes more complicated with increasing $n_c$ values
for degenerate cases. For this reason we use an alternative approach
where we vary the Debye frequency $\omega_D$ and the coupling
strength $\lambda$ in order to vary $\delta/\Delta$ ratio.
The motivation from physical point of view is the
expectation that number parity effect, if it exists,
should mainly be a function of dimensionless parameter $\delta/\Delta$.
Our numerical results support our alternative approach, as well.
Given a $\delta/\Delta$ ratio, if we repeat our calculations
for increasing values of $n_c$, we observe that $\Delta_p^{(i)}/\Delta$
ratios do not change very much.

Exact diagonalization approach of MFF \cite{mastellone} and BH \cite{berger}
for the non-degenerate case
showed that both ground state properties and excitation gap are
parity dependent and functions of the ratio of level spacing to the BCS gap,
$\delta/\Delta$. However systems addressed by MFF and BH are limited to relatively 
small number of levels, $n_c$, taken around Fermi level. Practically 
$n_c\sim 15$ is an upper limit for any exact diagonalization 
scheme (either Lancsoz method or other methods) because of large memory 
space requirements. In Fig. \ref{non-deg} we reproduce the dependence
of parity gap parameter, $\Delta_p^{(i)}$, upon level spacing $\delta$.
Here our aim in reproducing these results, which are obtained by 
using much larger values of $n_c$ (such as $n_c=500$) 
for the non-degenerate case, is to compare them with those 
of MFF and BH. 
When 
$n_c$ increases we obtain a different behavior of 
$\Delta_p^{(1)}$ which was not observed by
MFF and BH. As the number of levels $n_c$
increases, instead of a monotonic behavior, 
$\Delta_p^{(1)}/\Delta$ curve exhibits a minimum 
at $\delta/\Delta \sim 0.5$, which can not be observed with smaller 
$n_c$ values (see Fig. \ref{non-deg}(a)). On the other hand,
for the dependence of $\Delta_p^{(2)}$ upon $\delta$ we observe
the same behavior as MFF and BH, that is a minimum at
$\delta/\Delta\sim 1$ (see Fig. \ref{non-deg}(b)).
Von Delft and Braun \cite{vondelft3} 
show similar results for  $\Delta_p^{(2)}$ in the 
non-degenerate case in order to compare RS exact solution 
to earlier approaches to the problem of mesoscopic superconductivity.

Next we performed calculations for degenerate
cases ($d\ge 2$). In Fig. \ref{deg} we present the results for 
doubly degenerate case ($d=2$) which show that 
parity effect is still there. Similar results 
are obtained for higher degeneracies ($d>2$). Some important
conclusions can be drawn by comparing Fig. \ref{non-deg}
and Fig. \ref{deg}, immediately. Firstly, $\Delta_p^{(i)}$s repeat 
themselves with a periodicity of two for the non-degenerate
case, whereas they have a periodicity of four 
for the doubly degenerate case, that is the parity effect 
parameters repeat themselves with a period of $2d$. Secondly,
one can immediately observe that there is a remarkable difference
between the ratios of two different $\Delta_p^{(i)}$s for
the degenerate and non-degenerate case. For example, 
the ratio $\Delta_p^{(1)}/\Delta_p^{(2)}$ (Fig. \ref{non-deg}(a) and (b))
at a fixed $\delta/\Delta$ value ($\sim 10$), is about 5, 
while $\Delta_p^{(2)}/\Delta_p^{(3)}$ for doubly degenerate 
case (Fig. \ref{deg}(b) and (c)) is about 1.5 for the same 
value of $\delta/\Delta$. We will come to a rough estimation
of these values in a short while.

As it has been mentioned earlier, for certain energy spectra and more 
generally for lattice symmetries, energy levels are degenerate near 
the Fermi energy. 
In case of a d-fold degenerate single level with our notation
$n_c=0$, and the ground state energy (\ref{single})
measured from the Fermi level is given by
\begin{equation}
E_0^{(n)}=\frac{\lambda\delta}{d}[N^2-(d+\frac{1+(-1)^n}{2})N]
\label{onelevel}
\end{equation}
where $N$ is the integer part of $n/2$ as above. Parity effect appears in
the second term on the right hand side. Depending upon whether $n$ is odd or
even, the factor in front of $N$ becomes $d$ or $d+1$, respectively. This is
nothing but the blocking effect of the single electron.
It is clear that when $d$ becomes larger, this effect will become
less important. 
By substituting Eq. \ref{onelevel} into Eq. \ref{parity}
we obtain
\begin{equation}
\frac{\Delta_p^{(odd)}}{\Delta_p^{(even)}}=1+\frac{1}{d}
\label{ratio}
\end{equation}
where $odd$ ($even$) stands for $i$ odd (even) in Eq. \ref{parity}. Note 
that we consider a fixed chemical potential $\mu$ in derivation of  
(\ref{ratio}). Therefore this ratio (\ref{ratio})
is valid for a parity effect parameter $\Delta_p^{(i)}$
for which all ground state energies, $E_0^{(n)}$, 
used in Eq. \ref{parity} 
calculated with the same 
chemical potential $\mu$.  
Comparing Figs. \ref{deg}(b) and (c) at $\delta/\Delta\sim 1$
we find that $\Delta_p^{(3)}/\Delta_p^{(2)}\sim 1.33$, which is quite 
close to $1+1/2=1.50$.
Moreover, for $\delta/\Delta\sim 10$, we find 
$\Delta_p^{(3)}/\Delta_p^{(2)}\sim 1.51$ which is even closer to the 
ratio above. 
For 4-fold degeneracy, where $1+1/d=1.25$, 
we obtain similar ratios. For example
$\Delta_p^{(3)}/\Delta_p^{(4)} \sim 1.17$ at $\delta/\Delta \sim 1$,
and $\Delta_p^{(3)}/\Delta_p^{(4)} \sim 1.24$ at $\delta/\Delta \sim 10$.
These results are not unexpected 
since as $\delta/\Delta$ increases, contribution of electrons at the 
Fermi level to the ground state energy becomes dominant which lets single
level result become a better approximation. On the other hand when $\mu$ 
shifts, Eq. \ref{ratio} is not valid anymore and the ratio
$\Delta_p^{(odd)}/\Delta_p^{(even)}$ becomes very 
different from 1. For example, $\Delta_p^{(1)}/\Delta_p^{(2)}\sim 5$ for 
non-degenerate case and $\Delta_p^{(1)}/\Delta_p^{(4)}\sim 5$ for 
doubly degenerate case at $\delta/\Delta\sim 10$.

\section{Parity effect in Atomic Cluster}
We supplement the investigation of a parity effect ``from above"
(from macroscopic sizes down to mesoscopic sizes) by an investigation 
``from below", that is, starting from small clusters of atoms 
coupled at sites by some interaction energy \cite{dagotto}.
Unlike Eq. \ref{bcs}, we choose the negative-$U$ Hubbard interaction
\begin{equation}
H=-t\sum_{i\ne j}a_{i\sigma}^\dagger a_{j\sigma} + U \sum_{i=1}^N n_{i\uparrow}
n_{i\downarrow}
\label{hubbard}
\end{equation}
where $i$ and $j$  numerate atomic sites, $a_{i\sigma}^\dagger$ is 
the creation operator at
site $i$ with spin projection $\sigma$, and $n_{i\sigma}$ is the number operator.
This above Hamiltonian will reduce to a BCS Hamiltonian (\ref{bcs})
in the limit of large number of sites and small $U$ with the difference that 
the interaction (\ref{hubbard}) is a non-retarding
one, and the interaction shell (the cut-off energy) at 
Fermi energy is increased up to the value of the order of 
Fermi energy itself. In the absence of interactions for energy levels
of cubic cluster, we find $-3t$, $-t$, $t$ and $3t$ where $-t$ and $t$ are 
triply degenerate, and hence level spacing $\delta$ is given by $2t$.
An exact solution of Hamiltonian 
(\ref{hubbard}) in a cubic cluster allowing for maximum of 16 electrons 
has been studied earlier \cite{boyaci1}. The Hamiltonian matrix 
corresponding to Eq. \ref{hubbard} has a maximum dimension of 
$\sim6\times10^4$ after reduction of the size with symmetry 
consideration, and ground state eigenvalue is calculated by an exact
diagonalization method (of non-Lanczos type)\cite{boyaci2}.
Fermi operators $a_{i\sigma}$ are represented as
\begin{eqnarray}
a_{1\uparrow}=a\otimes u \otimes u \otimes u \otimes u \ldots \otimes u \nonumber \\
a_{1\downarrow}=v \otimes a \otimes u \otimes u \otimes u \ldots \otimes u \nonumber \\
a_{2\uparrow}=v \otimes v \otimes a \otimes u \otimes u  \ldots \otimes u \\
\ldots \nonumber \\
a_{8\downarrow}=v \otimes v \otimes v \otimes v \otimes v \ldots  \otimes a \nonumber
\end{eqnarray}
where $a,u,v$ are $2\times 2$ matrices
\begin{equation}
a= \left( \begin{array}{cc}  
0 \,\, 1 \\ 0 \,\, 0  
\end{array} \right) , \,
u= \left( \begin{array}{cc}
1 \,\, 0 \\ 0 \,\, 1 
\end{array} \right) , \, 
v= \left( \begin{array}{cc}
1 & 0 \\ 0 \, & -1
\end{array} \right)
\end{equation}
and $\otimes$ is the Kronecker matrix product symbol. By changing $U$, 
we change the bulk energy gap $\Delta$ which we calculate by using 
BCS gap equation. For $-1\ge U/t \ge -10$, approximating the density of
states by $1/12t$, we see that the coupling parameter $\lambda \simeq U/12t$
changes between 0.08 and 0.83. Therefore, at least for small $|U|$ values, 
we are in the weak-coupling BCS regime where approximate solution of the 
gap equation can be written as $\Delta=12t\exp(-12t/|U|)$. 
Non-zero parity effect parameter, $\Delta_p$, is directly seen from 
the energy versus electron number plot (see Fig. 4).
We observe that $E_0^{(n)}$ is not a monotonic function of $n$.
The ground state energy exhibits dips for even number of electrons.
This is an unambiguous indication of the parity effect,
which is a direct manifestation of Cooper pairing.

The inset in Fig. 4 shows that $\Delta_p^{(8)}=E_0^{(7)}-(E_0^{(6)}+E_0^{(8)})/2$
is non-zero as long as there is an attractive interaction at sites, i.e. 
$U<0$. We find, on contrary, that $\Delta_p=0$ at $U>0$ and hence conclude that 
repulsive Hubbard model does not lead to superconductivity.
Nevertheless in principle this result does not exclude 
the possibility of superconductivity in a larger system with $U>0$. 

Presence of parity effect for negative-$U$ Hubbard Hamiltonian is
important since it shows that dependence of superconducting
properties on whether there are even or odd number of electrons
is not specific to BCS Hamiltonian.
Although our rule that the
ratio of two successive parity effect parameters is $1+1/d$ does not work,
we clearly see that $\Delta_p^{(i)}$ values form groups 
in parallel to the degeneracy
of levels. For example, the first line in Table \ref{clratios} corresponds
to a jump from $-3$ state to $-1$ state, while the next five lines correspond to
energy levels $-1$. It is possible to understand different behavior 
of $\Delta_p^{(i)}$ for $i=3,9$ and 15 by introducing analogue of chemical 
potential given by $\mu=(E_{\rm HOMO}+E_{\rm LUMO})/2$ where HOMO and LUMO
stand for highest occupied and lowest unoccupied molecular orbital, respectively.
For $i=3,9$ and 15, $\mu$ is different in each $E_0^{(k)}$ value
used in calculations of $\Delta_p^{(i)}$. For example, in case of 
$\Delta_p^{(9)}=|E_0^{(8)}-(E_0^{(7)}+E_0^{(9)})/2|$, 
$\mu=0, -1$ and 1 while for $\Delta_p^{(11)}=|E_0^{(10)}-(E_0^{(9)}+E_0^{(11)})/2|$,
$\mu=1$ for all three ground states.
Figure \ref{fig5} shows three different parity effect parameters,
$\Delta_p^{(9)}$, $\Delta_p^{(10)}$ and $\Delta_p^{(11)}$ as a function 
of level spacing $\delta=2t$. It is remarkable that curves are very similar 
to those obtained for BCS Hamiltonian. Again, due to degeneracy, 
$\Delta_p^{(9)}$ which involves the jump $-1 \rightarrow 1$ exhibit a slightly
different behavior in comparison to $\Delta_p^{(10)}$ and $\Delta_p^{(11)}$.
As far as filling of molecular orbitals is 
concerned $i=9$ case is analog of Fig. \ref{deg}(a), while 
$i=10$ and 11 correspond to Fig. \ref{deg}(b) and (c), respectively.

\section{Conclusion}
In conclusion, we have studied the effect of degeneracy of discrete energy
levels on superconducting properties of a small metallic grain. We observe 
that parity effect parameter, which is a measure of the dependence of energy on 
whether the number of electrons in the grain is even or odd, exhibits
a behavior similar to the non-degenerate case for small $d$. 
We reproduced the behavior of parity effect parameter in the well 
studied non-degenerate case in order to compare our exact results
with previous work. In that case the parity effect parameter exhibits
a minimum instead of a monotonic behavior.
For d-fold degenerate states
it turns out that there are 2d different parity gaps and furthermore both
approximate analytic solutions and exact numerical results suggest that ratio
of two successive parity effect parameters is nearly $1+1/d$. Therefore 
careful measurements of parity parameters can be used to determine the 
degeneracy of energy levels. With increasing degeneracy, 
the pairing effect is suppressed.
Although there is no direct evidence for existence of SET transistors 
with a metallic grain of perfect crystal structure, observation
of perfect geometries in small grains prepared by vapor condensation 
technique opens a possibility for having highly symmetric structures.
As we discussed above smallness of the average level spacing in comparison to the 
level spacing due to finite size effect leads to a nearly degenerate
energy spectrum. Convergence of the ratio of two successive parity 
effect parameters to $1+1/d$ as the level spacing increases, can be used to detect
the degeneracy.

We also show that parity effect is not specific to BCS Hamiltonian by 
solving negative-$U$ Hubbard model for a small 
atomic cluster exactly. We can say that this is the most 
rigorous way to treat the problem since it does not involve any approximation 
like pairing assumption. Our results clearly show that parity 
effect exists and hence we conclude that it is model independent and
a general property of small Fermi systems with attractive interaction.
Furthermore, $\Delta_p^{(i)}/\Delta$ curves exhibit a behavior 
which is very similar to BCS case.
Finally, grouping of $\Delta_p^{(i)}$ values according to energy levels
of atomic cluster shows that degeneracy still manifests itself.

This work was partially supported by the Scientific and Technical Research
Council of Turkey (TUBITAK) under grant No. TBAG 1736

\begin{table}
\caption{
$\Delta_p^{(i)}=\left|E_0^{(i-1)}-\left(E_0^{(i-2)}+E_0^{(i)}\right)/2\right|$,
for different values of $i$ with $U=-1$. We observe three relatively 
large $\Delta_p^{(i)}$ values ($i=3,9,15$) corresponding to jumps between 
the energy levels of the cubic cluster ($-3 \rightarrow -1$, $-1 \rightarrow 1$,
and $1 \rightarrow 3$).}
\label{clratios}
\end{table}

\begin{center}
\begin{tabular}{|c|c|}
 $i$ & $\Delta_p^{(i)}$ \\ \hline
 3   & 1.003 \\        
 4   & 0.204 \\       
 5   & 0.212 \\      
 6   & 0.201 \\      
 7   & 0.206 \\
 8   & 0.197 \\
 9   & 1.010 \\
 10  & 0.197 \\
 11  & 0.206 \\
 12  & 0.201 \\
 13  & 0.212 \\
 14  & 0.204 \\
 15  & 1.003 \\     
\end{tabular}
\end{center}

\begin{figure}
\caption{Behavior of roots of Richardson-Sherman formula
in the doubly degenerate case. (a) For even number of bosons, $N$,
all roots are complex conjugate pairs, and real parts,
$x_\gamma$, are plotted with respect to $\lambda$.
Since the roots, $E_{2\gamma}$ and $E_{2\gamma-1}$
are complex conjugate pairs, their imaginary parts ($y_\gamma$)
cancel each other and the ground state energy 
becomes twice the sum of all $x_\gamma$. 
(b) When the
number of bosons, $N$, is odd, $N-1$ roots $E_q$ are complex 
conjugate pairs and one remaining root ($E_{21}$, shown by a darker solid line
in (b)) is real for all $\lambda$. Again only the real parts of
roots, $x_\gamma$, are plotted with respect $\lambda$.}
\label{roots}
\end{figure} 

\begin{figure}
\caption{Parity effect parameters as a function of level spacing for
non-degenerate case. (a) and (b) $n_c=15,30,60,120,240,360,500$ 
from bottom to top.
(a) With larger values of $n_c$, which are more realistic in comparison
to experiments, we obtain a behavior of $\Delta_p^{(1)}$
which was not observed previously. Instead of monotonically
decreasing towards $\Delta_p^{(1)}/\Delta=1$, the curve
makes a minimum at $\delta/\Delta \sim 0.5$. The inset 
shows the details around the minima. However 
it still approaches to one after this minimum as expected.
(b) $\Delta_p^{(2)}$ has the same expected behavior as shown 
by many previous studies. It has a minimum at $\delta/\Delta \sim 1$ and 
after this minimum it turns upward towards $\Delta_p^{(2)}/\Delta=1$}
\label{non-deg}
\end{figure}

\begin{figure}
\caption{Parity effect parameters as a function of level spacing for
doubly degenerate energy levels. $n_c=20,25,30$ from bottom to top in (a), 
(b) and (d), from top to bottom in (c). 
The immediate fact observed by comparing this figure with 
the one for the non-degenerate case exhibits itself when 
the ratios of different $\Delta_p^{(i)}$s are considered.
For example, when we take ratio of $\Delta_p^{(2)}$ to 
$\Delta_p^{(3)}$ (compare (b) and (c))we see that its value is much 
smaller than such a ratio in the non-degenerate case (compare Fig. 2(a) and (b)).}
\label{deg}
\end{figure}

\begin{figure}
\caption{Dependence of ground state energy $E_0^{(n)}$ (in units of $t$) 
upon number of particles $n$ for 
$U/t=-2,-3,-4$ from top to bottom.
The ground state energy 
exhibits drops for even number of electrons.
Lower inset shows the cubic cluster 
on which negative-$U$ Hubbard model is solved. 
Parity effect parameter $\Delta_p$ versus
on-site interaction $U$ at half filling is shown in the upper inset.}
\end{figure}

\begin{figure}
\caption{Parity effect parameters 
$\Delta_p^{(9)}$, $\Delta_p^{(10)}$ and $\Delta_p^{(11)}$ 
as a function of level spacing $\delta=2t$, for cubic cluster
where (a) $\Delta_p^{(9)}=|E_0^{(8)}-(E_0^{(7)}+E_0^{(9)})/2|$
(b) $\Delta_p^{(10)}=|E_0^{(9)}-(E_0^{(8)}+E_0^{(10)})/2|$ and 
$\Delta_p^{(11)}=|E_0^{(10)}-(E_0^{(9)}+E_0^{(11)})/2|$. 
The insets show the corresponding energy levels of the cluster. 
Since $\Delta_p^{(9)}$ involves the jump $-1\rightarrow 1$, 
it is relatively large in comparison 
to $\Delta_p^{(10)}$ and $\Delta_p^{(11)}$ as it can be seen in Table I. }
\label{fig5}
\end{figure}

\newpage
Fig. 1

\psfig{file=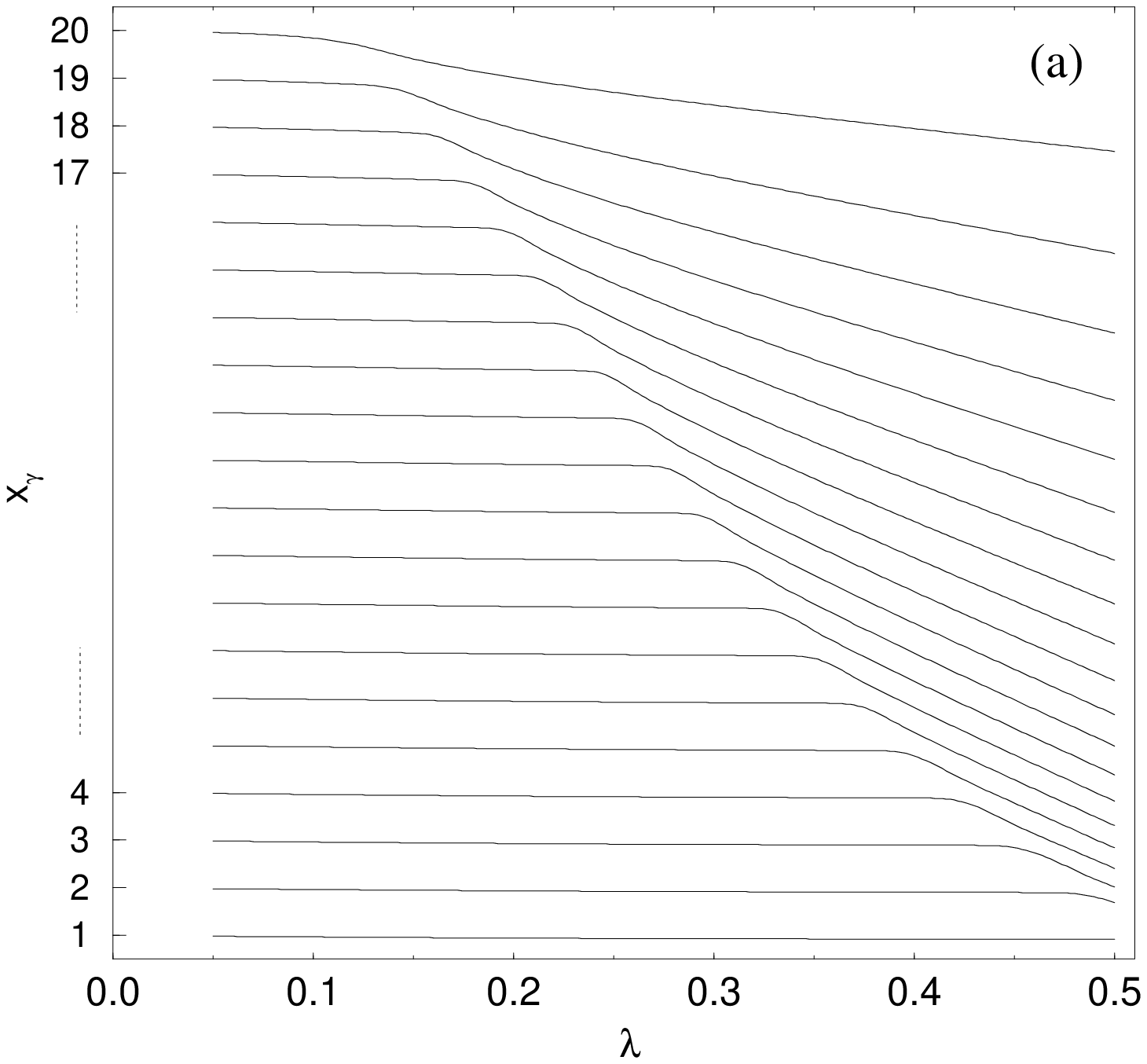,height=9cm,width=9cm}

\psfig{file=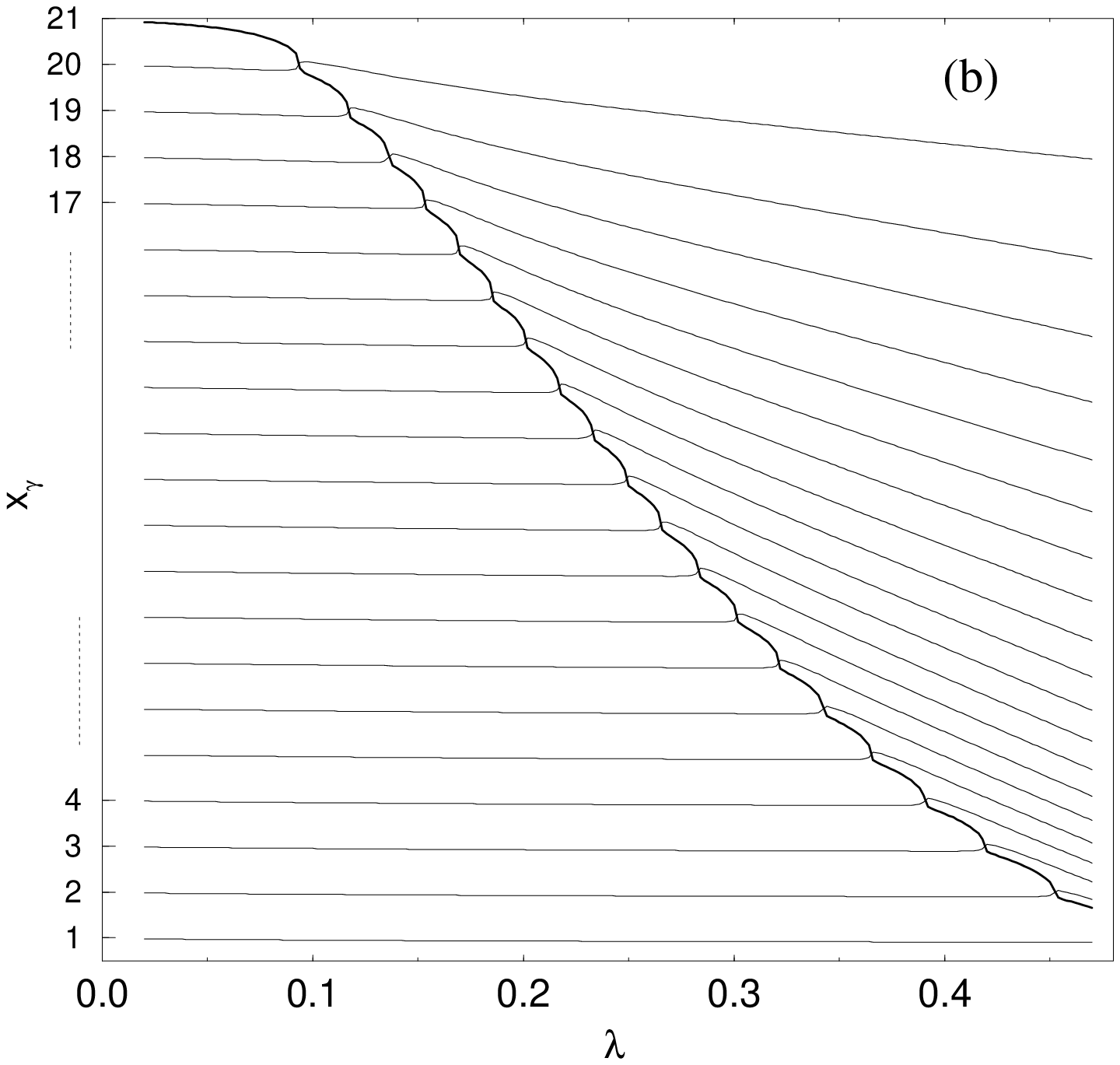,height=9cm,width=9cm}
\newpage
Fig. 2

\psfig{file=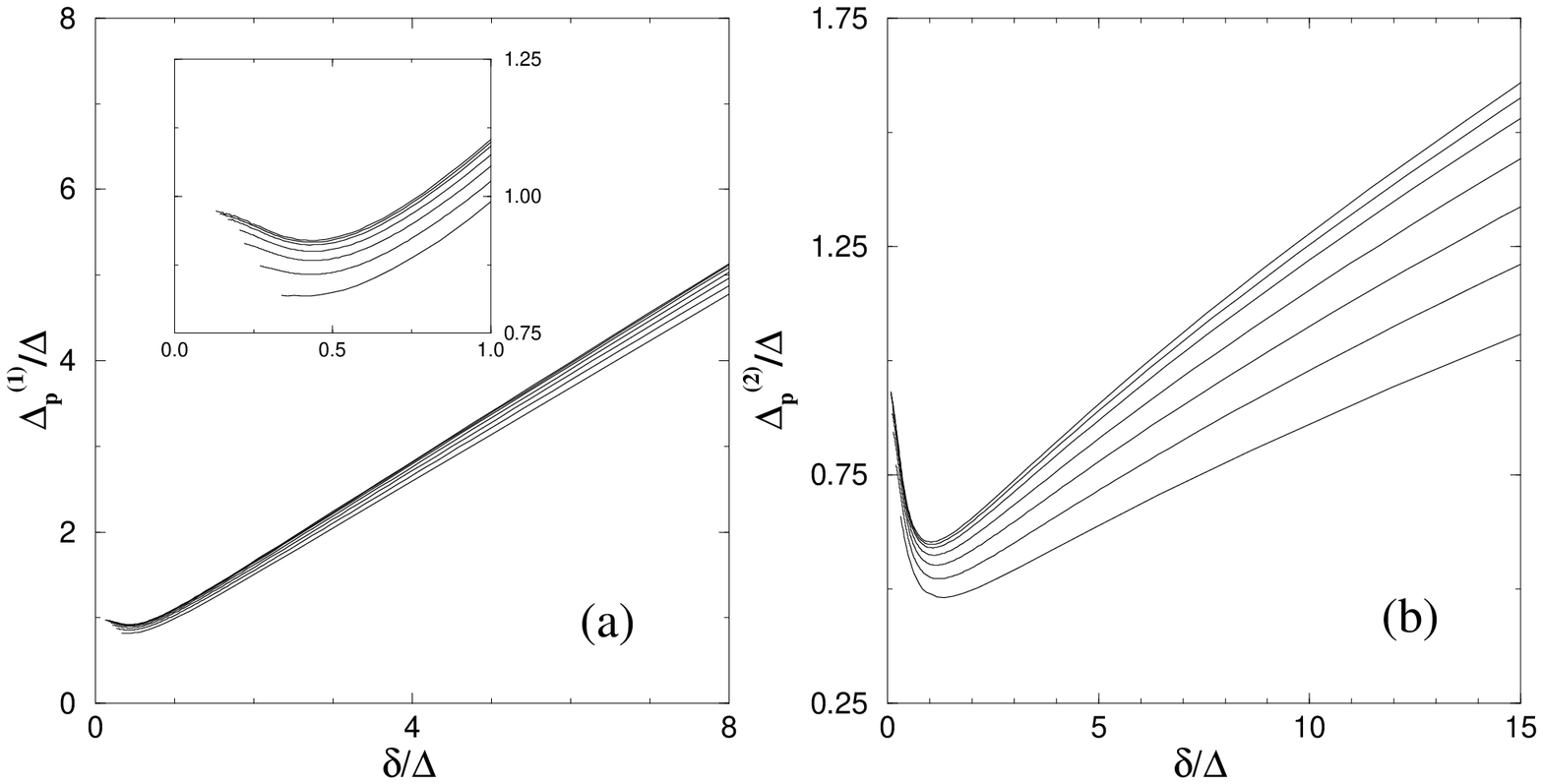,height=8cm,rwidth=16cm}
\newpage
Fig. 3

\psfig{file=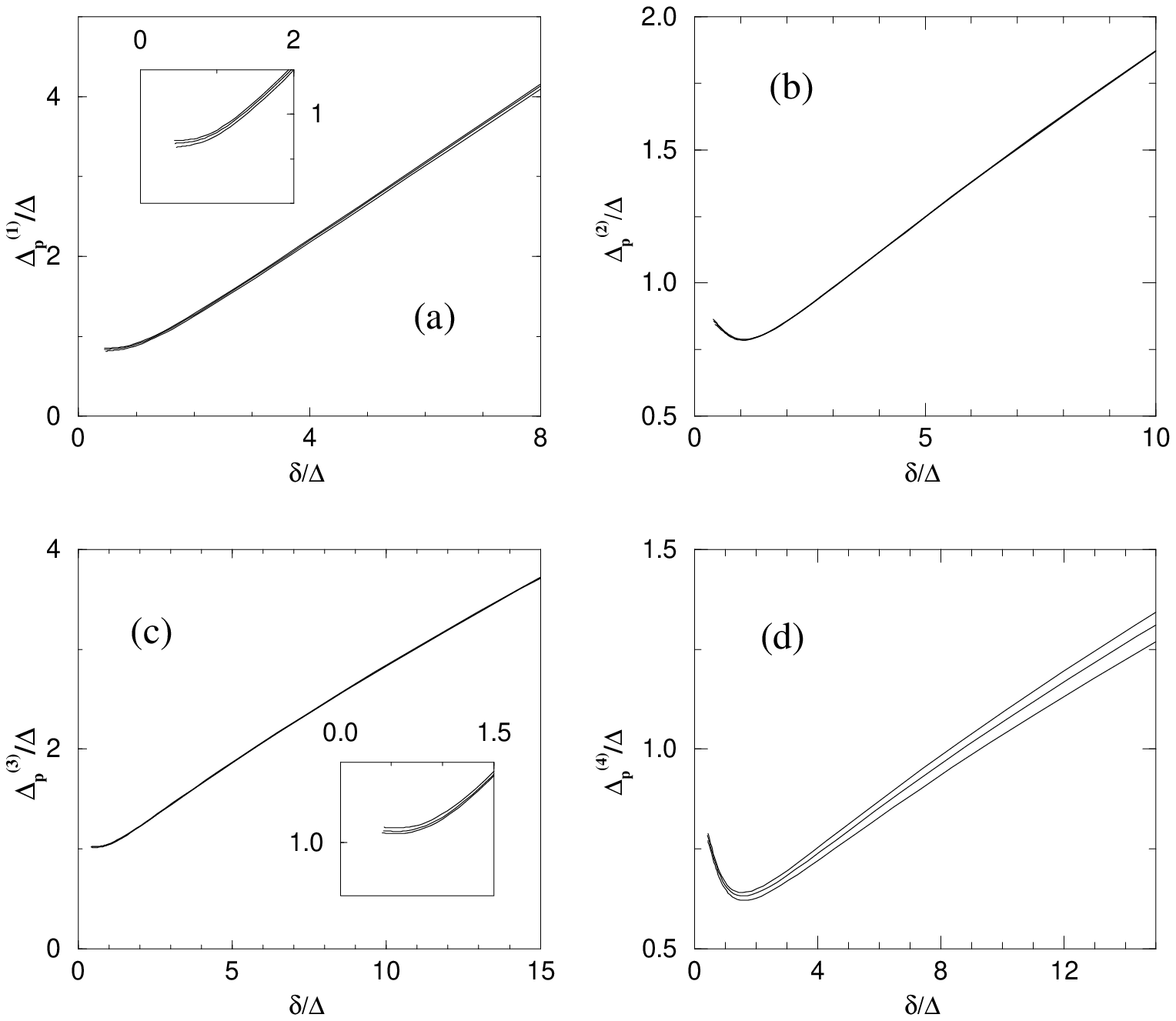,rheight=1cm,rwidth=1cm}
\newpage
Fig. 4

\hspace{-3cm} \psfig{file=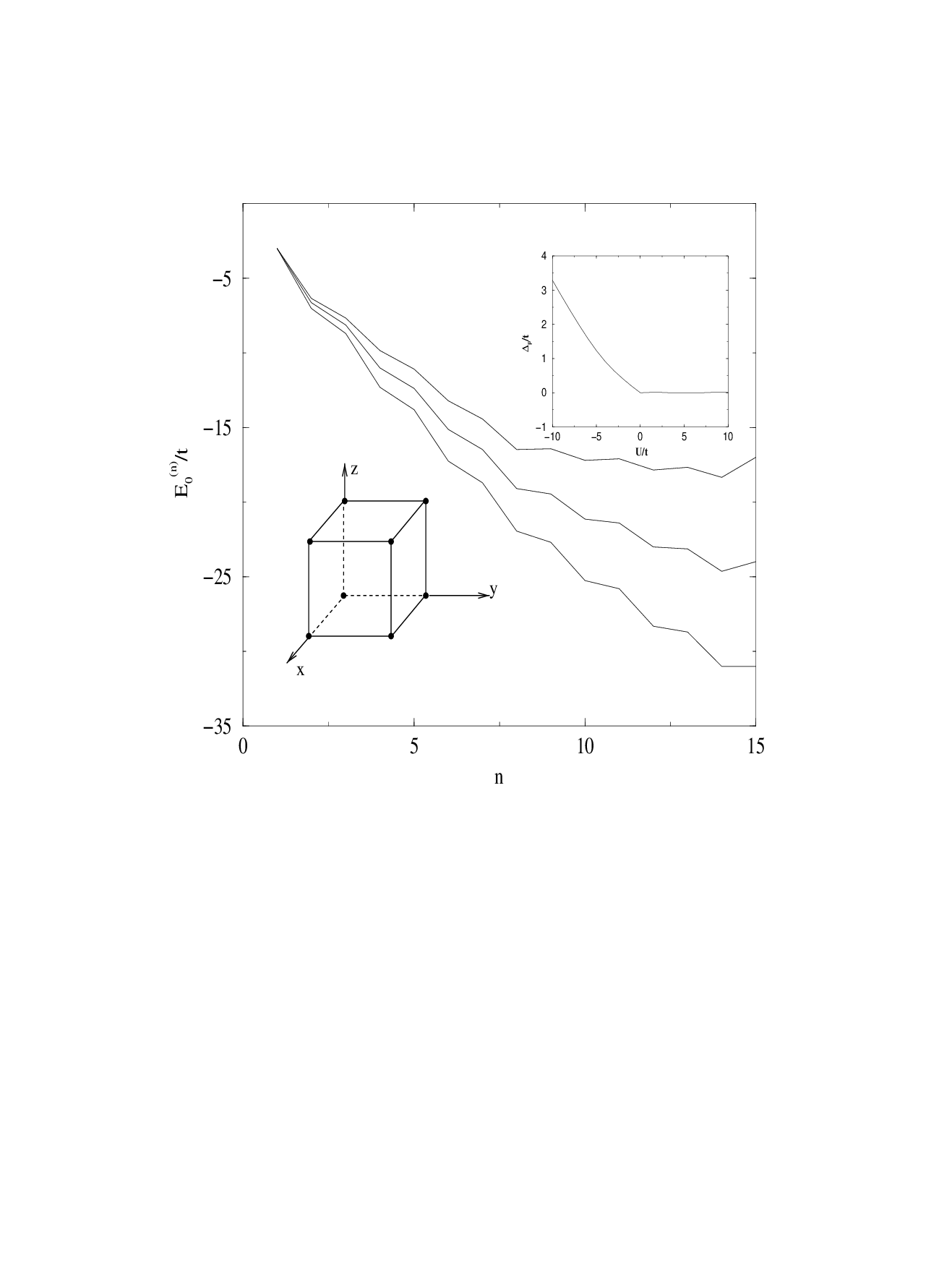,rheight=1cm,rwidth=1cm}

\newpage
Fig. 5

\psfig{file=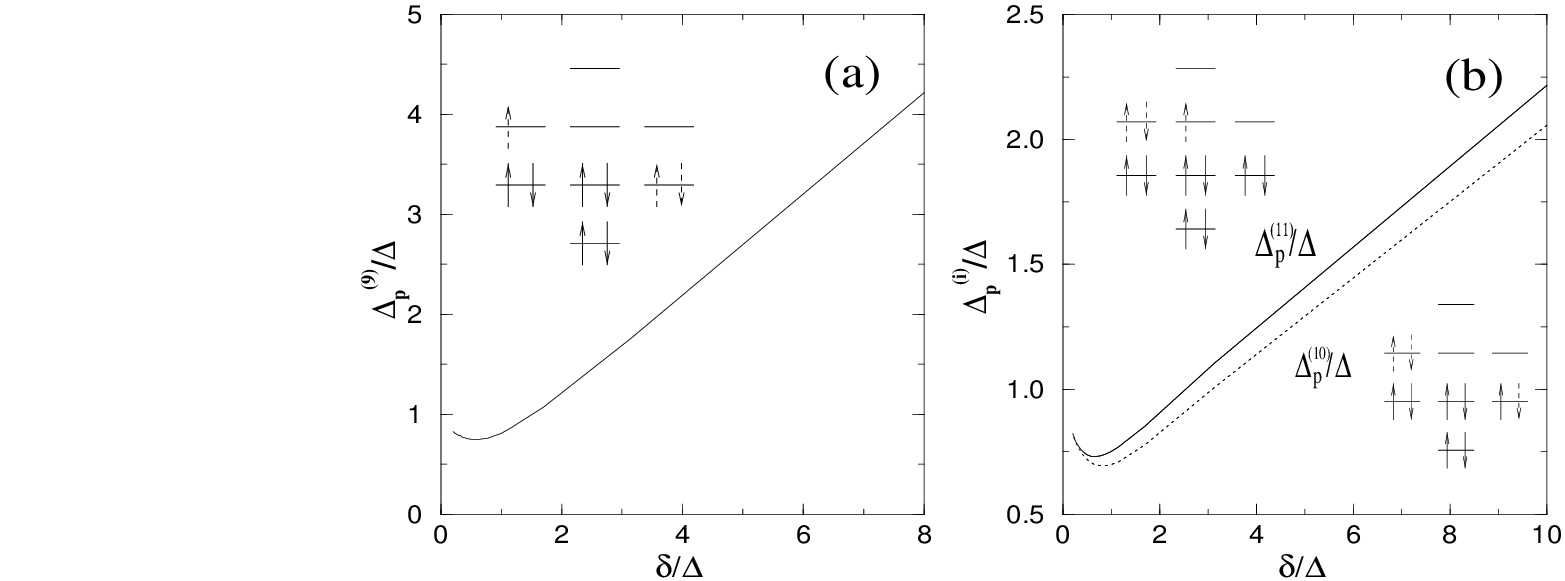,height=22cm,width=16cm}

\end{document}